\documentstyle [12pt]{article}
\newcommand{\agt}{{\lower 2pt\hbox{$>$} \atop \raise 2pt\hbox{$\sim$}}}
\newcommand{\alt}{{\lower 2pt\hbox{$<$} \atop \raise 2pt\hbox{$\sim$}}}

\begin{document}
\rightline{CMU-HEP94-02}
\rightline{DOE-ER/40682-55}
\leftline{hep-ph/9404271}
\leftline{March, 1994}
\vspace{5cm}
\begin{center}
{\LARGE\bf ORIGIN AND MECHANISMS OF CP VIOLATION}
\end{center}
\bigskip
\begin{center}
{\large\bf YUE-LIANG WU} \\
 Department of Physics, \\ Carnegie-Mellon University, \\
Pittsburgh, Pennsylvania 15213, U.S.A.
\end{center}
\newpage
\rightline{CMU-HEP94-02}
\rightline{DOE-ER/40682-55}
\leftline{hep-ph/9404271}
\leftline{March, 1994}
\bigskip
\begin{center}
{\Large\bf Origin and Mechanisms of CP Violation}
\end{center}
\bigskip
\begin{center}
{\bf Yue-Liang Wu \\
\small Department of Physics, Carnegie-Mellon University, \\
\small Pittsburgh, Pennsylvania 15213, U.S.A.}
\end{center}
\bigskip
\date{March, 1994}
\bigskip

\begin{abstract}
It is shown  that the mechanism of spontaneous symmetry breaking provides
not only a mechanism for giving mass to the bosons and the fermions, but also
a mechanism for generating CP-phase of the bosons and the fermions. A
two-Higgs doublet model  with Vacuum CP Violation and Approximate Global
$U(1)$ Family Symmetries (AGUFS) is built and may provide one of the simplest
and attractive models in understanding  origin and mechanisms of CP violation
at the weak scale.  It is seen that CP violation occurs everywhere it can from
a single CP phase of the vacuum and is generally classified into four types of
CP-violating mechanism. A new type of CP-violating mechanism is emphasized and
can provide a consistent application to both the established and the reported
CP-, T- and P-violating phenomena.
\end{abstract}

\bigskip


\newpage

In the Standard Model with three families, CP violation is known to
occur through a single physical KM-phase\cite{KM} in the gauge interactions of
the quarks. This phase originally comes from the complex Yukawa couplings.
Beyond the Standard Model, CP violation can take place through superweak
interactions \cite{LW}. It was pointed out by T.D. Lee \cite{TDL} that
CP symmetry could be broken spontaneously, thus the scalar particles are
responsible for a CP violation.  In the Weinberg three-Higgs doublet model
\cite{SW1}, CP violation appears through Scalar-Pseudoscalar Mixings (SPM).
Recently, I investigated \cite{YLWU1} a simple two-Higgs Doublet Model (2HDM)
and observed that  there exists a new type of CP-violating mechanism.
 By this new mechanism, the CP-violating parameters $\epsilon$ and
$\epsilon'/\epsilon$ in kaon decay and the neutral electric dipole
moment can be consistently accommodated.

 Before having definitive experimental test on origin and
mechanisms of CP violation, we should consider all the possibilities.
In this paper, I shall explicitly show how four types of
CP-violating mechanism can be induced from a single CP-violating phase of
the vacuum in a simple 2HDM.

  Such a 2HDM within the framework of the $SU(2) \times
U(1)$ gauge theory is built by considering the two basic
assumptions: first, CP violation solely originates from a single CP-violating
phase of the vaccum; second, the theory possesses Approximate Global U(1)
Family Symmetries (AGUFS) which act only on the fermions. In general,
without imposing any additional conditions, the AGUFS will naturally lead to
a Partial Conservation of Neutral Flavor (PCNF).

 It is known that the first assupmtion can be realized by the Spontaneous
CP Violation (SCPV), nevertheless, SCPV encounters so-called domain-wall
problem at the weak scale.  In order to prevent the domain-wall problem
from arising explicitly, we observe the following fact that

   {\it In the gauge theories of spontaneous symmetry breaking (SSB),
CP violation can be required solely originating from the vacuum after SSB,
even if CP symmetry is not good prior to the symmetry breaking.} The
{\it demanded condition} for such a statement is:
{\it CP nonconservation occurs only at one place of the interactions in the
Higgs potential}.  This requirement actually results that
the vacuum must violate CP symmetry. In particular, this condition may be
simply realized by an {\it universal rule}. That is, in a renormalizable
lagrangian all the  interactions with dimension-four  conserve CP and
only  interactions with dimension-two possess CP nonconservation.
It may also be naturally implemented through imposing some symmetries.
For convenience of mention, we refer such a CP violation  as a
Vacuum CP Violation (VCPV).

 For a 2HDM,  the most interesting case is the one with the universal rule
stated above. Then the Higgs potential can be simply written in the following
general form
\begin{eqnarray}
V(\phi) & = & \lambda_{1}(\phi_{1}^{\dagger}\phi_{1}- \frac{1}{2} v_{1}^{2})^2
+ \lambda_{2}(\phi_{2}^{\dagger}\phi_{2} - \frac{1}{2} v_{1}^{2})^2 \nonumber
\\
& & + \lambda_{3}(\phi_{1}^{\dagger}\phi_{1} - \frac{1}{2} v_{1}^{2})
(\phi_{2}^{\dagger}\phi_{2} - \frac{1}{2} v_{2}^{2})
+ \lambda_{4}[(\phi_{1}^{\dagger}\phi_{1})(\phi_{2}^{\dagger}\phi_{2})
-(\phi_{1}^{\dagger}\phi_{2})(\phi_{2}^{\dagger}\phi_{1})]  \nonumber  \\
& &  + \frac{1}{2}\lambda_{5}(\phi_{1}^{\dagger}\phi_{2} +
\phi_{2}^{\dagger}\phi_{1} - v_{1}v_2 \cos\delta )^2
+ \lambda_{6}(\phi_{1}^{\dagger}\phi_{2}- \phi_{2}^{\dagger}\phi_{1} -
v_{1}v_{2} \sin\delta )^{2}  \\
& & + [\lambda_{7}(\phi_{1}^{\dagger}\phi_{1} - \frac{1}{2} v_{1}^{2})
+ \lambda_{8} (\phi_{2}^{\dagger}\phi_{2} - \frac{1}{2} v_{2}^{2})]
[\phi_{1}^{\dagger}\phi_{2} + \phi_{2}^{\dagger}\phi_{1} -
v_{1}v_2\cos\delta ]  \nonumber
\end{eqnarray}
where the $\lambda_i$ ($i=1, \cdots, 8$) are all real parameters.
If all the $\lambda_i$ are non-negative,  the minimum
of the potential then occurs at $<\phi_{1}^{0} > = v_1 e^{i\delta}/\sqrt{2} $
and $<\phi_{2}^{0} > = v_2/\sqrt{2}$. It is clear that in the above potential
CP nonconservation can only occur through the vacuum, namely $\delta \neq 0$.
Obviously, such a CP violation appears as an explicit one in the potential
when $\lambda_{6} \neq 0$, so that the domain-wall problem does not explicitly
arise. Note that in general one can also demand one of other terms, such as
$\lambda_{5}$ or $\lambda_{7}$ or $\lambda_{8}$ to be complex in a general
potential.

To see the second assumption of the AGUFS, we start with a general Yukawa
interaction

\begin{equation}
L_{Y} = \bar{q}_{L}\Gamma^{a}_{D} D_{R}\phi_{a} + \bar{q}_{L}\Gamma^{a}_{U}
U_{R}\bar{\phi}_{a} + \bar{l}_{L}\Gamma^{a}_{E} E_{R}\phi_{a} + H.C.
\end{equation}
where $q^{i}_{L}$, $l^{i}_{L}$ and $\phi_{a}$ are $SU(2)_{L}$ doublet quarks,
leptons and Higgs bosons, while $U^{i}_{R}$, $D^{i}_{R}$ and $E^{i}_{R}$
are $SU(2)_{L}$ singlets. $i = 1,\cdots , n_{F}$ is a family label and
$a = 1, \cdots , n_{H}$ is a Higgs doublet label. $\Gamma^{a}_{F}$
($F= U, D, E$) are the arbitrary real Yukawa coupling matrices.

 It is known that in the limit that CKM matrix is unity, any models
with NFC \cite{GWP} at tree level generate global $U(1)$ family symmetries,
i.e., under the global U(1) transformations for each family of the fermions
 $(U,\ D )_{i}\rightarrow e^{i\alpha_{i}} (U, \ D)_{i} $,
the lagrangian is invariant. Where $\alpha_{i}$ are the constants and depend
only on the family index. In the realistic case, it is known that CKM
matrix deviates only slightly from unity. This implies that at the
electroweak scale any successful models can only possess approximate
global $U(1)$ family symmetries.

 With this consideration, we are motivated to parameterize the Yukawa coupling
matrices in such a convenient way that violations of the global
$U(1)$ family symmetries for the charged currents and the neutral
currents can be easily distinguished and the magnitudes of their violations
are characterized by the different sets of parameters.  This can be
implemented explicitly by parameterizing the matrices $\Gamma^{a}_{F}$ in
terms of the following general structure
\begin{equation}
\Gamma^{a}_{F}  =  O_{L}^{F} \sum_{i,j=1}^{n_{F}}\{ \omega_{i} (g_{a}^{F_{i}}
\delta_{ij}  + \zeta_{F} \sqrt{g^{F_{i}}} S_{a}^{F}\sqrt{g^{F_{j}}} )
\omega_{j} \} (O_{R}^{F})^{T}
\end{equation}
with $g^{F_{i}} = | \sum_{a} g^{F_{i}}_{a} \hat{v}_{a} |
/ (\sum_{a} |\hat{v}_{a}|^{2})^{\frac{1}{2}}$ and $\{\omega_{i}, i=1,
\cdots,n_{F}\}$ the set of diagonalized projection matrices
$(\omega_{i})_{jj'} = \delta_{ji}\delta_{j'i}$. $\hat{v}_{a}\equiv
<\phi^{0}_{a}>$ ($a=1, \cdots , n_{H}$) are Vacuum Expectation Values (VEV's)
which will develop from the Higgs bosons after SSB. $g^{F_{i}}_{a}$ are the
arbitrary real Yukawa coupling constants. By convention, we choose
$S_{a}^{F}=0$ for $a=n_{H}$ to eliminate the non-independent parameters.
$S_{a}^{F}$ ($a\neq n_{H}$) are the arbitrary off-diagonal real matrices.
$g^{F_{i}}$ are  introduced so that a comparison between the diagonal and
off-diagonal matrix elements becomes  available. $\zeta_{F}$ is a conventional
parameter introduced to scale the off-diagonal matrix elements with
the normalization $(S_{1}^{F})_{12}\equiv 1$ and $(S_{a}^{F})_{ij}$ being
expected to be of order unity (some elements of $S_{a}^{F}$ may be off
by a factor of 2 or more).  $O_{L,R}^{F}$ are the arbitrary orthogonal
matrices. Note that the above parameterization is general but
very useful and powerful for our purposes in analysing various
interesting physical phenomena.

 In general, one can always choose, by a redifinition of the  fermions,
a basis so that $O_{L}^{F}=O_{R}^{F}\equiv O^{F}$ and $O^{U} = 1$ or
$O^{D} = 1$ as well as $O^{E}=1$ since the neutrinos are considered to be
massless in this model. In this basis, the AGUFS and PCNF then imply that
\begin{equation}
(O^{F})_{ij}^{2} \ll 1 \  , \qquad i\neq j \ ; \qquad  \zeta_{F}^{2}  \ll 1 \
{}.
\end{equation}
where $O^{F}$ describe the AGUFS in the charged currents and $\zeta_{F}$
mainly characterizes the PCNF. Obviously, if taking $\zeta_{F} =0$ , it turns
to the case  of NFC at tree level. Furthermore, when
$\zeta_{F} =0$ and $O^{D} = O^{U} = 1$, the theory possesses global
$U(1)$ family symmetries.

We then conclude that the smallness of the CKM mixing angles and the
suppression of the FCNSI can be attributed to the AGUFS and PCNF and can be
regarded as being naturally in the sense of the 't Hooft's
criterion \cite{THOOFT}.
It is actually manifest since  exact global U(1) family symmteries demand
that all the off-diagonal interactions should disappear from the model.

 Let us now present a detailed description for the model with VCPV and
AGUFS.  The physical interactions  are usually given in the mass basis of
the particles. For the simplest 2HDM, the physical basis  after SSB is defined
through $f_L = (O_{L}^{F}V_{L}^{f})^{\dagger}F_L$ and $ f_R = (O_{R}^{F}P^f
V_{R}^{f})^{\dagger}F_R $
with $V_{L,R}^{f}$ being unitary matrices and introduced to diagonalize
the mass matrices
\begin{equation}
(V_{L}^{f})^{\dagger}(\sum_{i}m_{f_{i}}^{o}\omega_{i} + \zeta_{F} c_{\beta}
\sum_{i,j} \sqrt{m_{f_{i}}^{o}} \omega_{i} S_{1}^{F} \omega_{j}
\sqrt{m_{f_{j}}^{o}} e^{i\sigma_{f}(\delta - \delta_{f_{j}})}) V_{R}^{f} =
\sum_{i} m_{f_{i}}\omega_{i}
\end{equation}
with $m_{f_{i}}$ the masses of the physical states $f_{i}= u_i, d_i, e_i$.
Where $m_{f_{i}}^{o}$ and $\delta_{f_{i}}$ are defined via
\begin{equation}
(c_{\beta} g_{1}^{F_{i}}e^{i\sigma_{f}\delta} +
s_{\beta} g_{2}^{F_{i}})v \equiv \sqrt{2}m_{f_{i}}^{o} e^{i\sigma_{f}
\delta_{f_{i}}}
\end{equation}
 with $v^2 = v_{1}^{2} + v_{2}^{2}= (\sqrt{2}G_{F})^{-1}$,
$ c_{\beta}\equiv \cos\beta = v_1/v$ and $s_{\beta}\equiv \sin\beta
= v_2/v$. $P^{f}_{ij} = e^{i\sigma_{f} \delta_{f_{i}}}\delta_{ij}$,
with $\sigma_{f} =+$, for $f= d, e$, and  $\sigma_{f} = - $, for $f = u$.

  For convenience of discussions, it is simple to make the
phase convention by writting $V_{L,R}^{f} \equiv  1 + \zeta_{F} T_{L,R}^{f}$.
 In a good approximation, to the first order in $\zeta_{F}$ and the lowest
order in $m_{f_{i}}/m_{f_{j}}$ with $i < j$, we find
that $m_{f_{i}}^{2} \simeq (m_{f_{i}}^{o})^{2} + O(\zeta^{2}_{F})$ and
for $i < j$
\begin{eqnarray}
(T_{L}^{f})_{ij} & \simeq & - (T_{L}^{f})_{ji}^{\ast} \simeq c_{\beta}
\sqrt{\frac{m_{f_{i}}}{m_{f_{j}}}} (S_{1}^{F})_{ij} e^{-i\sigma_{f} (\delta
- \delta_{f_{j}})} + O((\frac{m_{f_{i}}}{m_{f_{j}}})^{3/2}, \zeta_{F}) \\
(T_{R}^{f})_{ij} & \simeq & - (T_{R}^{f})_{ji}^{\ast} \simeq c_{\beta}
\sqrt{\frac{m_{f_{i}}}{m_{f_{j}}}} (S_{1}^{F})_{ji} e^{-i\sigma_{f} (\delta
- \delta_{f_{j}})} + O((\frac{m_{f_{i}}}{m_{f_{j}}})^{3/2}, \zeta_{F})
\end{eqnarray}

 With this phase convention, the CKM matrix $V$ has the following form
\begin{equation}
V   = (V_{L}^{U})^{\dagger} (O_{L}^{U})^{T} O_{L}^{D} V_{L}^{D} \equiv
V^{o} + V'
\end{equation}
where $V^{o} \equiv  (O_{L}^{U})^{T} O_{L}^{D}$ is a real matrix and
$V' \simeq \zeta_{D}[V^{o} T_{L}^{d}] + \zeta_{U}[V^{o}T_{L}^{u}]^{\dagger}$
is a complex matrix.

{\it  The scalar interactions of the fermions} read in the physical basis
\begin{eqnarray}
L_{Y}^{I} & = & (2\sqrt{2}G_{F})^{1/2}\sum_{i,j,j'}^{3}\{
\bar{u}_{L}^{i} V_{ij'} (m_{d_{j'}}\xi_{d_{j'}}\delta_{j'j} + \zeta_{D}
\sqrt{m_{d_{j'}}m_{d_{j}}} S_{j'j}^{d}) d^{j}_{R}H^+
- \bar{d}_{L}^{i} V_{ij'}^{\dagger} (m_{u_{j'}}\xi_{u_{j'}}\delta_{j'j}
\nonumber \\
& & + \zeta_{U}\sqrt{m_{u_{j'}}m_{u_{j}}} S_{j'j}^{u}) u^{j}_{R}H^-
+ \bar{\nu}_{L}^{i} (m_{e_{i}}\xi_{e_{i}}\delta_{ij} + \zeta_{E}
\sqrt{m_{e_{i}}m_{e_{j}}} S_{ij}^{e}) e^{j}_{R}H^+  + H.C. \} \nonumber \\
& & + (\sqrt{2}G_{F})^{1/2}\sum_{i,j}^{3} \sum_{k}^{3}\{ \bar{u}^{i}_{L}
(m_{u_{i}} \eta_{u_{i}}^{(k)} \delta_{ij} + \zeta_{U}\sqrt{m_{u_{i}}m_{u_{j}}}
S_{k,ij}^{u}) u^{j}_{R} + \bar{d}^{i}_{L}
(m_{d_{i}} \eta_{d_{i}}^{(k)} \delta_{ij}  \\
 & & +\zeta_{D}\sqrt{m_{d_{i}}m_{d_{j}}} S_{k,ij}^{d}) d^{j}_{R}
+ \bar{e}^{i}_{L}(m_{e_{i}} \eta_{e_{i}}^{(k)} \delta_{ij} +
\zeta_{E}\sqrt{m_{e_{i}}m_{e_{j}}} S_{k,ij}^{e}) e^{j}_{R}
+ H.C. \} H_{k}^{0} \nonumber
\end{eqnarray}
with (in the above good approximations)
\begin{eqnarray}
& & \xi_{f_{i}} \simeq \frac{\sin\delta_{f_{i}}}
{s_{\beta}c_{\beta}\sin\delta}e^{i \sigma_{f}(\delta - \delta_{f_{i}})}
 - \cot\beta \  ,  \\
& & S^{f}_{ij} \simeq s_{\beta}^{-1} (e^{i\sigma_{f}(\delta - \delta_{f_{j}})}
- \frac{\sin\delta_{f_{j}}}{\sin\delta} ) (S_{1}^{F})_{ij} \  , \qquad
S^{f}_{ji} \simeq  s_{\beta}^{-1} (e^{i\sigma_{f}(\delta - \delta_{f_{i}})}
- \frac{\sin\delta_{f_{j}}}{\sin\delta} ) (S_{1}^{F})_{ji} \  . \\
& & \eta_{f_{i}}^{(k)}  = O_{2k}^{H}  +  (O_{1k}^{H} + i \sigma_{f} O_{3k}^{H})
\xi_{f_{i}} \ ; \qquad S_{k,ij}^{f} = (O_{1k}^{H} + i \sigma_{f}
O_{3k}^{H} )S^{f}_{ij}\ .
\end{eqnarray}
where $O_{ij}^{H}$ is the orthogonal matrix introduced to redefine the
three neutral scalars $ \hat{H}_{k}^{0} \equiv (R, \hat{H}^{0}, I)$
into their mass eigenstates $H_{k}^{0} \equiv (h, H, A)$, i.e. $\hat{H}_{k}^{0}
 =  O^{H}_{kl} H_{l}^{0}$ with $(R+iI)/\sqrt{2} =
s_{\beta}\phi^{0}_{1}e^{-i\delta} - c_{\beta} \phi^{0}_{2}$ and
$(v+\hat{H}^{0} +iG^{0})/\sqrt{2} = c_{\beta}\phi^{0}_{1}
e^{-i\delta} + s_{\beta} \phi^{0}_{2} $.
Here $H_{2}^{0}\equiv H^{0}$ plays the role of the Higgs boson in the
standard model. $ H^{\pm}$ are the charged scalar pair with $ H^{\pm} =
s_{\beta}\phi^{\pm}_{1}e^{-i\delta} - c_{\beta} \phi^{\pm}_{2}$.

  By explicitly giving the new interactions, we now briefly discuss
in this short paper their main features and summarize the most interesting
physical phenomena arising from these interactions. The systematic analyses
and detailed calculations are presented in a longer paper\cite{YLWU2}.

 1) \  We observe that if the relative phase between the two VEV's is
nonzero,  each fermion ($f_{i}$) and neutral scalar $H_{k}^{0}$  are then
characterized not only by their physical mass $m_{f_{i}}$ and
$m_{H_{k}^{0}}$   but also by a physical phase $\delta_{f_{i}}$ and
$ \delta_{H_{k}^{0}} \equiv arg(O_{1k}^{H} + i \sigma_{f} O_{3k}^{H})$
respectively. This shows that the Higgs mechanism
provides not only a mechanism for gaving mass to the bosons and the
fermions, but also a mechanism for generating CP-phase of the bosons and the
fermions.

 2) \ All the vacuum-induced CP violations can be classified  into four
types of mechanism according to their origins and/or interactions.
To be more clear, we emphasize as follows

 {\bf Type-I.} \  The new type of CP-violating mechanism \cite{YLWU1,YLWU2}
which arises from the induced complex diagonal Yukawa couplings $\xi_{f_{i}}$.
Such a CP violation can occur through both charged- and neutral-scalar
exchanges.

{\bf Type-II.}\  Flavor-Changing SuperWeak (FCSW)-type  mechanism.
This type of mechanism also occurs through both charged- and neutral-scalar
exchanges and is described by the complex coupling matrices $S_{ij}^{f}$
in this model.

 {\bf Type-III.}\  The induced KM-type CP-violating mechanism  which
is characterized in this model by the complex parameters $\zeta_{F} T_{L}^{f}$
and occurs in the charged gauge boson and  charged scalar interactions of the
quarks.

 {\bf Type-IV.}\  The Scalar-Pseudoscalar Mixing (SPM) mechanism
which is described by the mixing matrix $O_{kl}^{H}$ and the phases
$arg(O_{1k}^{H}+i\sigma_{f}O_{3k}^{H})$. This type of CP
violation appears in the purely bosonic interactions and also
in the neutral-scalar-fermion interactions in this
model. In general, SPM mechanism can also occur in the charged-scalar-fermion
interactions when there exist more than two charged scalars, for example,
the Weinberg 3HDM.

 3) \  Without making any  additional assumptions, $m_{f_{i}}$, $V_{ij}$,
$\delta_{f_{i}}$ (or $\xi_{f_{i}}$), $\delta$, $\tan\beta$, $\zeta_{F}$,
$(S_{1}^{F})_{ij}$ (or $S_{ij}^{f}$), $m_{H_{k}^{0}}$, $m_{H^{+}}$ and
$O_{kl}^{H}$ are in principle all the free parameters and will be determined
only by the experiments. Nevertheless, from the AGUFS and PCNF,  we can  draw
the general features that $V_{ij}^{2} \ll 1$ for $i \neq j$ and
$\zeta_{F}^{2} \ll 1$. The $m_{f_{i}}$ and $V_{ij}$ already appear in the SM
and have been extensively investigated. For the other parameters, it is
expected that  $(S_{1}^{F})_{ij}$ are of order unity.
Moreover, in order to have the FCNSI be suppressed manifestly, it is in favor
of having  $\tan\beta > 1$ and  $|\sin\delta_{f_{j}}/\sin\delta | \alt 1 $
(see eq.(12)). The diagonal scalar-fermion Yukawa
couplings $\eta_{f_{i}}^{(k)}$ or $\xi_{f_{i}}$ can be, for the light
fermions,  much larger than those in the SM  and may, of course, also be
smaller than those in the SM (the latter case appears to happen for heavy
top quark). Nevertheless, the former case is more
attractive since it will result in significant interesting phenomenological
effects.  The most interesting choice for large $\xi_{f_{i}}$ is
$\tan\beta \gg 1$ since it favors the suppression of the FCNSI.

4)\ From the established $K^{0}-\bar{K}^{0}$ and $B^{0}-\bar{B}^{0}$ mixings,
we obtain that $\zeta_{D}/s_{\beta} < 10^{-3} m_{H_{k}^{0}}/GeV$.
{}From the current experimental bound of the $D^{0}-\bar{D}^{0}$ mixing,
i.e. $\Delta m_{D} < 1.3 \times 10^{-4}$ eV, we have
$\zeta_{U}/s_{\beta} < 3\times 10^{-3} m_{H_{k}^{0}}/GeV$. Note that
this can only be regarded as an order-of-magnitude estimation since in
obtaining
these values we have used the vacuum insertion approximation for the
evaluation of the hadronic matrix elements. From the established CP-violating
parameter $\epsilon$, it requires either to fine-tune the parameters
$\delta_{d}$ , $\delta_{s}$, $arg(O_{1k}^{H} + i O_{3k}^{H})$,
$(S_{1}^{D})_{12}$ {\it et al}, so that the effective CP-phase is of order
$10^{-2}$ or to choose $\zeta_{D}/s_{\beta} \alt 10^{-4} m_{H_{k}^{0}}/GeV$.
For the latter case, the CP-violating phases are indeed generically of order
unity.

 5) \  To see how the various mechanisms play the role on CP violation and
provide interesting physical phenomena, let us consider the
following three cases:

(i) when $\zeta_{F}/s_{\beta}\ll 1$ with  $(S_{1}^{F})_{ij}\sim O(1)$,
 $m_{H^{+}} < v =246$ GeV and  $|\xi_{i}| \gg 1$ ($i\neq t$), it is obvious
that only the new type of CP-violating mechanism (type-I) plays an
important role.  The effects from the FCSW-type and KM-type mechanisms
are negligible.

 In this case, the CP-violating parameter $\epsilon$ can be
fitted from the contributions of the long-distance dispersive
effects\cite{DP,DH,HYC}  through the $\pi$, $\eta$ and $\eta'$ and/or
short-distance box graph with charged scalar exchanges. This is easily
implemented in our model through choosing appropriate parameter
$Im(\xi_{s}\xi_{c})$ and/or $Im(\xi_{s}\xi_{c})^{2}$, respectively, for a
given mass $m_{H^{+}}$.  The ratio $\epsilon'/\epsilon$ is expected to be of
order $10^{-3}$ from the long-distance contribution which was first
correctly estimated by Donoghue and Holstein\cite{DH} and has generally been
discussed in \cite{HYC}, and from the tree level diagram with charged scalar
exchange. The latter case is easily reached by choosing
appropriate parameters $Im(\xi_{s}\xi_{d}^{\ast})$ and $Im(\xi_{s}\xi_{u})$.
The neutron EDM $d_{n}$ can be consistently accommodated by choosing other
independent parameters, such as $Im(\xi_{d}\xi_{c})$ and $Im (\eta_{d}^{(k)}-
\eta_{u}^{(k)})^{2}$ for the one-loop contribution with charged scalar and
neutral scalar exchanges respectively, and $Im(\xi_{t}\xi_{b})$ and
$Im(\eta_{t}^{(k)})^{2}$ for the Weinberg gluonic operator contribution
with charged scalar and neutral scalar exchanges respectively, as well as
$Im(\xi_{t}\xi_{q})$ ($q=d,s,u$) for the quark gluonic chromo-EDM.
The electon EDM $d_{e}$ from Barr-Zee two-loop mechanism \cite{BZ} is expected
to be in the present experimental sensitivity for appropriate values of
$Im(\eta_{t}^{(k)}\eta_{e}^{(k)})$ and $Im(\eta_{t}^{(k)}\eta_{e}^{(k) \ast})$
as well as $O_{2k}^{H} Im\eta_{e}^{(k)}$.  CP violation in the hyperon
decay can also be significant in this case. Based on the general analyses in
\cite{DHP}, we have, for example, the CP asymmetry observable $A(\Sigma^{\pm}
\rightarrow n\pi^{\pm}) \sim 10^{-3}$. Direct CP violation
in B-meson decay is, however, small in this limit case. Nevertheless, T-odd and
CP-odd triple-product correlations could be substantial in the inclusive
and exclusive semileptonic decays of B-meson into the $\tau$ leptons.

 In addition, by including the new contributions to the neutral meson mixings
from the box diagrams with charged-scalar exchange,
the mass difference $\Delta m_{K}$ can be reproduced by a purely
short-distance analysis when $|\xi_{c}| \gg 1$.
For example, for $B_{K} = 0.7$, and $m_{H^{+}} = 100$ GeV, it needs
$|\xi_{c}|\simeq 13$. When $|\xi_{t}|\sim 1$, $B^{0}-\bar{B}^{0}$ and
$B_{s}^{0}-\bar{B}_{s}^{0}$ mixings can also receive a contribution as large as
the one in the standard model.

(ii) when $10^{-4}m_{H_{k}^{0}}/GeV \alt \zeta_{D}/s_{\beta} < 0.1$,
$\zeta_{U}/s_{\beta} < 0.3$ for $m_{t} = 150$ GeV and
$(S_{1}^{F})_{ij}\sim O(1)$,, $|\xi_{i}| \sim 1$, both the new type of
CP-violating mechanism and the induced KM-type mechanism become less important
and the parameter $\epsilon$ is then accounted for by the FCSW-type mechanism
(type-II) together with the SPM mechanism (type-IV). If the CP-violating
phases are indeed generically of order unity, thus the ratio
$\epsilon'/\epsilon$ becomes unobservable small ($\sim 10^{-6}$).
In this case, its effects in the B-system are also small.

(iii) when $\zeta_{D}/s_{\beta} \sim 0.2$ and $\zeta_{U}/s_{\beta} \sim 0.6$
for $m_{t}=150$ GeV, $c_{\beta}\sim s_{\beta}$, $|\xi_{i}| \sim 1$ and
$m_{H^{0}_{k}} \gg v =246$ GeV, i.e., neutral scalars are very heavy, then
the CP-violating mechanism is governed by the induced KM-type mechanism.
But it could be still different from the standard KM-model if the charged
scalar is not so heavy, this is because the new contributions from diagrams
with charged-scalar exchange can be significant. Therefore only when
the charged scalar also become very heavy, the induced KM-type mechanism
then approaches to the standard KM-model which have been extensively
studied \cite{YLWU3}.

  It is seen that precisely measuring the direct CP violation in kaon
(and hyperon) decays and the direct CP violation in B-system as well as
the EDM's of the electron and the neutron are very important for clarifying
origin and mechanisms of CP violation. For instance, if the direct CP
violation in kaon decay is big and of order $10^{-3}$, the
electron EDM is also in the present observable level, while direct CP
violation in B-system is unobservable small, we then
conclude that the new type of CP-violating mechanism will be important.

  Based on the assumption of the AGUFS and PCNF, the mass of the scalars
could be less constrained from the indirect experimental data. Searching for
these exotic scalars  is worthwhile at both $e^{+}e^{-}$ and hadron colliders.
It is believed that the mechanisms of CP violation discussed in this
model should also play an important role in understanding the baryogenesis
at the electroweak scale\cite{CKN}. In particular, its requirement for
relatively light Higgs bosons is in favor of our model. We may conclude
that if one Higgs doublet is necessary for the generation of
the mass of the bosons and the fermions, then two Higgs doublets are needed
for origin and phenomenology of CP violation and also for baryogenesis at
the electroweak scale.

  Finally, I would like to remark that as this model is the simplest extension
of the Standard Model, we do not expect to be able to answer the questions
which also appear in the standard model, such as many free parameters and
the hierarchic properties of the parameters. We have also restricted ourselves
in this paper to the weak CP violation and have ignored the strong CP problem.
What we have shown is that such a simple
2HDM with VCPV and AGUFS possesses very rich phenomenological features,
in particular on the phenomenology of CP violation. What we may do is to
determine and/or restrict all the physical parameters from the direct and/or
indirect experimental measurements, just like what we have been doing
for the Standard Model.

  I benefited from discussions with R.F. Holman, L.F. Li, E.A. Paschos
and L. Wolfenstein. I also wish to thank Prof. K.C. Chou for having introduced
me to this subject.  This work was supported by DOE
grant \# DE-FG02-91ER40682.


\newpage

\begin{thebibliography}{99}
\bibitem{KM} M. Kobayashi and T. Maskawa, Prog. Theor. Phys. {\bf 49}, 652
(1973).
\bibitem{LW} L. Wolfenstein, Phys. Rev. Lett. {\bf 13}, 562 (1964).
\bibitem{TDL} T.D. Lee, Phys. Rev. {\bf D8}, 1226 (1973); Phys. Rep. {\bf 9},
143 (1974).
\bibitem{SW1} S. Weinberg, Phys. Rev. Lett. {\bf 37}, 657 (1976).
\bibitem{YLWU1} Y.L. Wu,  CMU Report, CMU-HEP93-11, 1993; CMU-HEP93-28,
DOE-ER/40682-53, hep-ph/9312348, 1993.
\bibitem{GWP} S.L. Glashow and S. Weinberg, Phys. Rev. {\bf D15}, 1958 (1977);
E.A. Paschos, Phys. Rev. {\bf D15}, 1966 (1977).
\bibitem{THOOFT} G. 't Hooft, in {\it Recent Developments in Gauge Theories},
Cargese Summer Institute Lectures, 1979, edited by G. 't Hooft {\it et al.}.
\bibitem{YLWU2} Y.L. Wu,  CMU Report, CMU-HEP94-01, hep-ph/9404241,
 80 pages, 1994.
\bibitem{DP}  Y. Dupont and T.N. Pham, Phys. Rev. {\bf D28}, 2169 (1983);
D. Chang, Phys. Rev. {\bf D25} (1982) 1318; J. Haglin, Phys. Lett.
{\bf B117}, 441 (1982).
\bibitem{DH} J.F. Donoghue and B.R. Holstein, Phys. Rev. {\bf D32}, 1152
(1985).
\bibitem{HYC} H.Y. Cheng, Phys. Rev. {\bf D34}, 1397 (1986).
\bibitem{DHP} J.F. Donoghue, X.G. He and S. Pakvasa, Phys. Rev. {\bf D34},
833 (1986).
\bibitem{YLWU3} See, for example, Y.L. Wu,  {\it Recent theoretical development
on direct CP violation $\epsilon'/\epsilon$}, in: Proc. of the XXVI Int..
Conf. on High Energy Phyics, p. 506, 1992-Dallas, Texas, edited by
J.R. Stanford, and references therein;
\bibitem{BZ} S.M. Barr and A. Zee, Phys. Rev. Lett. {\bf 65}, 21 (1990).
\bibitem{CKN} For a review see A.G. Cohen, D.B. Kaplan, and A. E. Nelson,
Ann. Rev. Nucl. Part. Sci. {\bf 43}, 27 (1993).
\end{thebibliography}
\end{document}